\documentclass[10pt,a4paper,oneside]{article}
\usepackage{amssymb}
\usepackage{amsmath}
\usepackage{graphicx}
\usepackage{theorem}
\usepackage{setspace}
\usepackage{appendix}

\usepackage{booktabs}
\usepackage{longtable}
\usepackage{rotating}
\usepackage{multirow}
\usepackage{threeparttable}
\usepackage[font=small,labelfont=bf]{caption}
\usepackage{enumerate}
\usepackage{url}
\usepackage{color}
\usepackage{lineno,hyperref}
\makeatletter

\newcommand{\Rmnum}[1]{\expandafter\@slowromancap\romannumeral #1@}
\makeatother

\usepackage{setspace}
\begin{document}
\title{{\LARGE \textbf{Fault diagnosis of rolling element bearings with a spectrum searching method}}}
\author{Wei Li$^{1}$, Mingquan Qiu$^{1}$, Zhencai Zhu$^{1}$, Fan Jiang$^{1}$ and Gongbo Zhou$^{1}$\\
\\
$^{1}$School of Mechatronic Engineering,\\
China University of Mining and Technology,\\
Xuzhou, 221116, P.R. China. Email: liwei\_cmee@163.com\\
\\
}

\date{}
\maketitle

\begin{abstract}
Rolling element bearing faults in rotating systems are observed as impulses in the vibration signals, which are usually buried in noises. In order to effectively detect the fault of bearings, a novel spectrum searching method is proposed. The structural information of spectrum (SIOS) on a predefined frequency grid is constructed through a searching algorithm, such that the harmonics of impulses generated by faults can be clearly identified and analysed. Local peaks of the spectrum are projected onto certain components of the frequency grid, and then the SIOS can interpret the spectrum via the number and power of harmonics projected onto components of the frequency grid. Finally bearings can be diagnosed based on the SIOS by identifying its dominant or significant components. Mathematical formulation is developed to guarantee the correct construction of the SIOS through searching. The effectiveness of the proposed method is verified with simulated signals and experimental signals.

\bigskip

\textbf{Keywords:} rotating machinery, bearings, fault diagnosis, spectrum searching \bigskip
\end{abstract}


\section{Introduction}
Rotating machinery is widely used in many industrial fields. Fault diagnosis in rotating machinery is important for system maintenance and process automation. In practice, faulty bearings contribute to most of the failures in rotating machinery \cite{jalan2009,jar2006,liu2012,abd2000}. It is reported that about 40\% to 90\% of failures are related to rolling element bearing failures from large to small machines \cite{bianchini2011}. Particularly, either inner-race or outer-race flaws dominate most bearing failures. Periodic sharp impulses characterize these faults, and the characteristic frequencies can be theoretically computed. Nevertheless those impulses are of low power and usually buried by noises. These signals are also usually modulated by some high-frequency harmonic components and resulted in a series of harmonics of characteristic frequencies \cite{randall2011}.

In order to realized fault diagnosis of bearing, some features of vibration signals are extracted through time-domain methods, frequency-domain methods, and time-frequency methods \cite{jar2006}. Time-domain methods are directly based on the time waveform, e.g. peak amplitude, root-mean-square amplitude, variance, skewness, kurtosis, correlation dimension and fractal dimension. Frequency-domain methods are based on the transformed signal in frequency domain, i.e. Fourier spectrum, cepstrum analysis, envelope spectrum. Wavelet analysis, short time Fourier transform, Wigner-Ville distribution and Hilbert-Huang transform are the time-frequency methods \cite{,li2012,xie2012fast,li2006wigner,peng2005}, which investigate waveform signals in both time and frequency domain. When used to analyse noisy bearing vibration signals, most of those methods may produce unsatisfactory results and may not give useful information about the characteristic frequencies in order to identify the faults.

Spectral kurtosis was developed to identify the characteristic frequencies of bearings, where a filter was designed to get the signal with the maximum kurtosis in spectrum and then the envelope analysis was usually applied to show the characteristic frequencies \cite{guo2014,wang2012,tian2016}. The wavelet techniques were widely used to decompose the vibration signals in order to find the most useful filter for fault diagnosis \cite{tai2011intel,wang2013,li2013envelop,liu2014adapt}. The conventional band-pass filters were also applied whose parameters were optimized through genetic algorithms or adaptive algorithms \cite{wang2011,kang2015envel}. Those methods are relatively complex which involve complicate computations.

Indeed spectral kurtosis based methods are to find the resonant frequency band of vibration signals which contains a train of high-energy harmonics of characteristic frequency, and then transform the resonant frequency band to a low frequency band through envelope analysis. It is suggested that the characteristic frequency could be identified by finding harmonics with high energies in spectra. Reference \cite{randall2011} also pointed out that the impulses of bearing faults could be detected when a series of harmonics of the characteristic frequency are identified in the spectra.

This paper proposes a simple method to detect the faults of bearings by searching the harmonics in spectra. As discussed before, the impulses generated by bearing faults are usually modulated, and harmonics of the bearing characteristic frequency exist in the spectrum of vibration signals. Hence we simply search the local peaks in the spectrum of vibration signals. Then the local peaks related to harmonics of certain frequency is projected onto a predefined frequency grid, such that the so-called structure information of the spectrum (SIOS) of the spectrum is constructed. The SIOS includes two defined indexes, which provide the information about the number and the power of harmonics of certain frequency. The dominant and significant components in the SIOS are just corresponding to the characteristic frequency of bearings and therefore the bearing faults can be diagnosed.

The rest of the paper is organized as follows. Section 2 introduces the spectrum searching method and the SIOS. In section 3, the method is applied to diagnose bearing with a simulated signal, and then it is compared with a benchmark study. The discussion and concluding remarks are given in section 4 and 5.

\section{Spectrum searching method}
As illustrated in \cite{randall2011}, faults of rolling element bearings generate impulses, and excites frequency resonances of the whole structure between the bearing and the transducer. The low harmonics of the bearing characteristic frequencies are usually masked by other vibration components; while the harmonics can be found easier in a higher frequency range, but higher harmonics may smear over one another. In case of heavy noises the harmonics series usually can not be directly recognized in the spectrum. The harmonics do exist in the spectrum, but it is not possible to determine the bearing characteristic frequencies by measuring the spacing of the harmonic series. Hence we propose the spectrum searching method to identify the bearing characteristic frequencies. The main steps are as follows:
\begin{itemize}
  \item find the local peaks with locally larger amplitudes in the spectrum by searching over the whole frequency range,
  \item construct the SIOS on the pre-defined frequency grid by projecting components of the spectrum with local peaks onto components of the frequency grid, and
  \item identify the dominant or significant components on the frequency grid according to the SIOS.
\end{itemize}

In this section we will describe the first two steps. For convenience in the rest of the paper, $P(k)$ is defined as the single-sided power amplitude of the $k$th frequency component in the spectrum; $F(k)$ is defined as the frequency in $Hz$ corresponding to the $k$th frequency component; and $I(k)$ is defined to indicate whether the $k$th frequency component has a local peak or not. The resolution of the spectrum is denoted as $\Delta_s$. The sampling rate of the vibration signal is denoted as $F_s$.

\subsection{Find the peaks in the spectrum}
The spectrum consists of the power amplitude of each frequency component. In order to construct the SIOS, we define the local peaks of the spectrum as follows.

\begin{figure}[!t]
  \centering
  \includegraphics[width=0.85\textwidth]{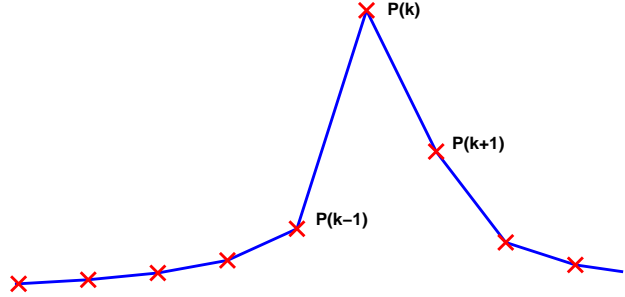}
  \caption{An example of local peaks.}
  \label{localpeak}
\end{figure}

\emph{Definition of Local Peaks.} Given three frequency components of the spectrum, i.e. $F(k-1)$, $F(k)$ and $F(k+1)$, then $P(k)$ is called a local peak if
\begin{equation}
P(k) > P(k-1)  \mbox{ and }  P(k)>P(k+1).
\label{peakdef}
\end{equation}
By searching on the spectrum, all frequency components satisfying inequality (\ref{peakdef}) can be found. Fig. \ref{localpeak} gives an example of local peaks.

Such a definition may lead to a large number of local peaks. Obviously we are not interested in the frequency components with small amplitudes which may be related to noises. The harmonics of bearing characteristic frequencies are usually with relatively larger amplitudes. Hence a threshold is proposed to suppress the influence of noises to some extent as follows
\begin{equation}
J_{th}(k) = \frac{1}{2l+1}\sum_{i=k-l}^{k+l}P(i) + \delta,
\label{localJ}
\end{equation}
where $l$ and $\delta$ are nonnegative constants. If $P(k)$ satisfies inequality (\ref{peakdef}) and
\begin{equation}
P(k)>J_{th}(k),
\label{thresh}
\end{equation}
then $I(k) = 1$, and otherwise $I(k) = 0$. The threshold in Eq. (\ref{localJ}) is varying in terms of $k$. The first part of the threshold is the moving average of power amplitudes; and the second part, i.e. $\delta$, is used to control the total number of identified local peaks.

\subsection{Construct the SIOS}
Assume the bearing characteristic frequencies are within the range from $F_l$ to $F_h$ in Hz ($F_l$ and $F_h$ are both frequency components of the spectrum), and the frequency grid is defined as
\begin{equation}
G = [F_l,F_l+\Delta_G,F_l+2\Delta_G,F_l+3\Delta_G,\cdots,F_h]
\label{basis}
\end{equation}
where $\Delta_G$ is a selectable positive constant. Therefore the number of frequency components of $G$ is $(F_h-F_l)/\Delta_G + 1$. We use $G(i)$ to represent the $i$th frequency component of $G$. $\Delta_G$ is the interval between components of the frequency grid, and it is suggested to be the resolution of the spectrum or its demultiplier, and we denote it as
\[
\Delta_G = \frac{\Delta_s}{\theta}, \theta \mbox{ is a positive integer and  }  \theta \geq 1.
\]

\begin{figure}[!t]
	\centering
	\includegraphics[width=0.85\textwidth]{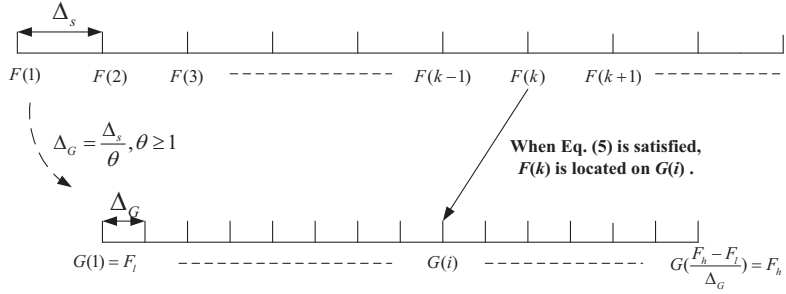}
	\caption{The relation between the spectrum and the frequency grid.}
	\label{grid}
\end{figure}

If a local peak is found on the $k$th frequency component of the spectrum, then it is projected onto the $i$th component of $G$ if
\begin{equation}
F(k)/G(i)  \mbox{  is an integer}
\label{nocal}
\end{equation}
with $i=1,2,\cdots,(F_h-F_l)/\Delta_G + 1$. In other word, we try to project all frequency components of the spectrum onto a frequency grid according to Eq. (\ref{nocal}). Fig. \ref{grid} illustrates the relation between the spectrum and the frequency grid.

Based on the frequency grid, we define two indexes to represent the SIOS. The first one is the number of local peaks projected onto the $i$th component of $G$, i.e. $N(i),i=1,2,\cdots,(F_h-F_l)/\Delta_G + 1$, and
\[
N(i) = \sum I(k), \mbox{where $G(i)$ and $F(k)$ satisfy Eq. (\ref{nocal})}.
\]
If $F(k)$ is projected onto a component of $G$ according to Eq. (\ref{nocal}), then the harmonics of $F(k)$  will also be projected onto the same component of $G$.

The second index is the total power of local peaks projected onto the $i$th component of $G$, i.e. $E(i),i=1,2,\cdots,(F_h-F_l)/\Delta_G + 1$, and
\[
E(i) = \sum P(k), \mbox{where $G(i)$ and $F(k)$ satisfy Eq. (\ref{nocal})}.
\]
This index is used to distinguish the useful signal from the noises. If $G(i)$ is corresponding to a characteristic frequency of bearings, $E(i)$ should be relatively large.

In case the characteristic frequency of bearings and its harmonics are multipliers of the $i$th frequency component of $G$, all of them will be projected onto the $i$th component of $G$. Then the corresponding $N(i)$ and $E(i)$ could be a dominant component in the SIOS. Nevertheless $\Delta_G$ and $\Delta_s$ can not be infinitely small. Hence the characteristic frequency of bearings and its harmonics may not be the frequency components of the spectrum, and they may also not be projected onto the component of $G$ according to Eq. (\ref{nocal}). For instance a given frequency, i.e. $f_m$, within the range of $G$ described by
\begin{equation*}
f_m = \alpha_{G(i)}\Delta_G+b_G, 0\leq b_G<\Delta_G,
\end{equation*}
\begin{equation}
F_l \leq \alpha_{G(i)}\Delta_G=G(i) \leq F_h
\label{fret}
\end{equation}
can not be projected onto $G$ when $b_G\neq 0$, where $\alpha_{G(i)}$ is a positive integer and the term $\alpha_{G(i)}\Delta_G$ is the $i$th component of $G$. The $j$th harmonic of $f_m$ denoted by $\beta_j f_m$, may also not be projected onto $G$, where $\beta_j=j$.

In order to overcome this problem, we reformulate Eq. (\ref{nocal}) as
\begin{equation}
\beta_j f_m/G(i) - \lfloor \beta_j f_m/G(i)\rfloor < \sigma(i,j),
\label{bound}
\end{equation}
where $\sigma(i,j)>0$ and $\lfloor \rfloor$ is the flooring operator. When inequality (\ref{bound}) is satisfied, we say the $j$th harmonic of $f_m$ is projected onto the $i$th component of $G$.

Now the key question is how to select $\sigma(i,j)$. Without loss of generality, $f_m$ can be assumed as
$F(k)\leq f_m <F(k+1)$ or $F(k-1)\leq f_m < F(k)$.

We would like to find $\sigma$ in inequality (\ref{bound}), such that $f_m$ and its harmonics are all projected onto the $i$th component of $G$ when inequality (\ref{peakdef}) and (\ref{bound}) are satisfied.

Considering the limitation of interval of $G$, we have
\[
\frac{\beta_j f_m}{\alpha_{G(i)}\Delta_G} = \frac{\beta_j(\alpha_{G(i)}\Delta_G+b_G)}{\alpha_{G(i)} \Delta_G} = \beta_j + \frac{\beta_j b_G}{\alpha_{G(i)} \Delta_G}.
\]
Since $0 \leq b_G < \Delta_G$, we have
\begin{equation}
\frac{\beta_j f_m}{\alpha_{G(i)}\Delta_G} = \beta_j +\frac{\beta_j b_G}{\alpha_{G(i)} \Delta_G} < \beta_j +\frac{\beta_j}{\alpha_{G(i)}}.
\label{basisc}
\end{equation}

Assume $\beta_j$ is smaller than $\alpha_{G(i)}$ (\textbf{Assumption 1}), then $0 < (\beta_j / \alpha_{G(i)}) < 1$. By setting
\begin{equation}
\sigma(i,j) = \frac{\beta_j}{\alpha_{G(i)}},
\label{sig}
\end{equation}
we have
\[
\frac{\beta_j f_m}{\alpha_{G(i)}\Delta_G} - \lfloor \frac{\beta_j f_m}{\alpha_{G(i)}\Delta_G}\rfloor < \sigma(i,j)
\]
according to inequality (\ref{basisc}). Clearly $f_m$ and its harmonics can be projected onto the same component on $G(i)$ when inequality (\ref{bound}) and Eq. (\ref{sig}) are satisfied.

In next, we will shown that the assumption 1 made on $\beta_j$ and $\alpha_{G(i)}$ can generally be satisfied. According to inequality (\ref{fret}), $\alpha_{G(i)}$ becomes larger when $\theta$ is larger, as
\begin{equation}
\frac{F_l}{\Delta_G} = \frac{\theta F_l}{\Delta_s}\leq \alpha_{G(i)} \leq \frac{\theta F_h}{\Delta_s} = \frac{F_h}{\Delta_G}.
\label{betajcons}
\end{equation}
Since $F_l\leq f_m \leq F_h$, $\beta_j$ in Eq. (\ref{sig}) is constrained by
\[
\beta_j F_l \leq \beta_j f_m \leq \frac{F_s}{2}
\]
where $F_s / 2$ is the maximum frequency of the spectrum. Then we have
\begin{equation}
\beta_j \leq  \frac{F_s}{2F_l}.
\label{betajc}
\end{equation}
We could take a sufficient large sampling length of vibration signal to obtain a fine spectrum resolution, such that
\begin{equation}
\Delta_s < \min_j\frac{F_l}{\beta_j}=\frac{2F_l F_l}{F_s}.
\label{gam}
\end{equation}
Then based on inequality (\ref{betajcons}) and inequality (\ref{gam}), we have
\[
\alpha_{G(i)} \geq \frac{\theta F_l}{\Delta_s} >\frac{\beta_j}{F_l} \theta F_l=\beta_j \theta
\]
and consequently $\beta_j \theta / \alpha_{G(i)}$.
That means the assumption 1 can be fulfilled if inequality (\ref{gam}) is satisfied.

With a defined $F_l$, inequality (\ref{gam}) can always be satisfied with a sufficient large sampling length of vibration signals. Hence the assumption 1 can be generally satisfied.

For instance, we are interested in the characteristic frequency in the range from $100$Hz to 200Hz, and therefore the frequency grid is defined with $F_l=100$Hz and $F_h=200$Hz. The sampling rate is 12kHz. According to in inequality (\ref{gam}), assumption 1 can be satisfied when
\[
\Delta_s < \frac{2F_l F_l}{F_s} = \frac{20000}{12000}=1.667.
\]
We take $2^{15}$ as the sampling length, and then $\Delta_s = 0.3662$ which is satisfying inequality (\ref{gam}). According to inequality (\ref{betajcons}) and (\ref{betajc}), we have
\[
\max_{i,j} \frac{\beta_j\theta}{\alpha_{G(i)}}=\frac{F_s\Delta_s}{2F_lF_l}=\frac{12000\times0.3662}{2\times100\times100}=0.2197<1,
\]
and therefore $\max \limits_{i,j} (\beta_j / \alpha_{G(i)})<1$, which means the assumption 1 is satisfied.

The SIOS, i.e. $N$ and $E$ on $G$, gives the information about the harmonics of the frequency components of $G$. The index $N(i)$ represents the number of harmonics of $G(i)$ found in the spectrum; and the index $E(i)$ represents the total power of harmonics of $G(i)$. In fact the spectrum is interpreted in terms of $G(i)$, $N(i)$ and $E(i)$.

The frequency component in SIOS is treated as a dominant one, when this component is significant in $N$ and $E$ at the same time. If $E(i)$ is relatively large and $N(i)$ is small, then $G(i)$ is much likely a discrete component of the spectrum without harmonics. If $N(i)$ is relatively large and $E(i)$ is small, then $G(i)$ is much likely corresponding to noises.

\begin{figure}[!t]
	\centering
	\includegraphics[width=0.85\textwidth]{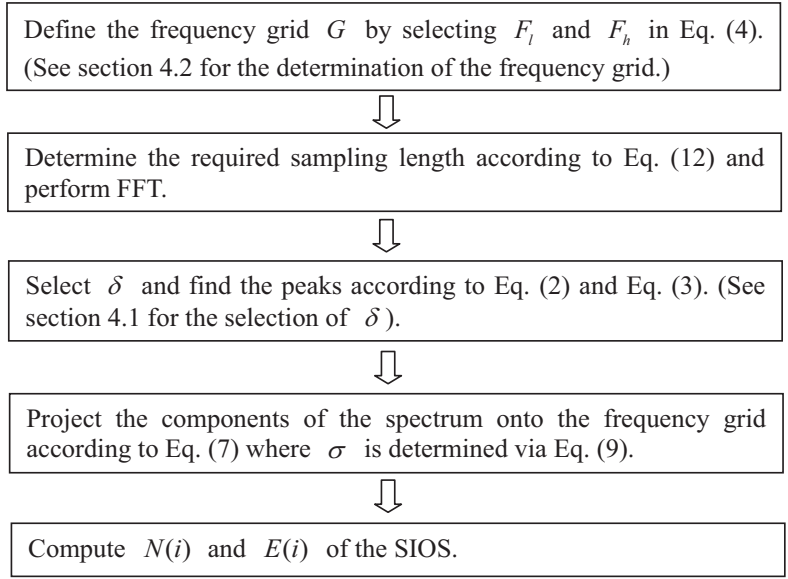}
	\caption{The flowchart for constructing SIOS.}
	\label{flow}
\end{figure}

The flowchart for constructing SIOS is given in Fig. \ref{flow}.

Moreover, let us suppose that the sampling frequency of a discrete time series $x(i), i=1, 2, \ldots, n$, satisfy the limiting condition in inequality (\ref{gam}), then the pseudo-code of SIOS algorithm for $x(i)$ is given in Table \ref{tab0}. Once the $N$ and $E$ are derived, they can be applied to perform bearing fault diagnosis. In next a simulation and two experimental studies are used to describe how to realize bearing diagnosis using the SIOS.
\begin{table}[!t]\normalsize
\renewcommand{\arraystretch}{1.3}
\begin{center}
\caption{The pseudo-code of SIOS algorithm.}\label{tab0}
\begin{tabular}{ll}
  \hline
  & \textbf{Inputs:} $F_l$, $F_h$, $P$, $F$, $F_s$, $\theta$, $\delta$, $l$ \\
  & \textbf{Outputs:} $N$, $E$ \\

  1& $L_F \leftarrow length(F)$, $\Delta_G \leftarrow \Delta_s/\theta$; \\
  2& $\Delta_s \leftarrow F_s/L_F$, $L_G \leftarrow (F_h - F_l)/\Delta_G$, $I(j) \leftarrow 0, j = 1, 2, \ldots, L_F$; \\
  3& $G(k) \leftarrow (F_l + k \times \Delta_G)$, $N(k) \leftarrow 0$ and $E(k) \leftarrow 0, k = 1, 2, \ldots, L_G$; \\

  & \textbf{Search local peaks:} \\
  4& \textbf{for} $k = (l + 1)$ to $(L_F - l)$ \\
  5& \qquad $ J_{th} \leftarrow \frac{1}{2l+1}\sum_{i=k-l}^{k+l}P(i) + \delta$; \\
  6& \qquad \textbf{if} $P(k) > P(k-1)$ and $P(k) > P(k+1)$ and $P(k) > J_{th}$ \textbf{then} \\
  7& \qquad \qquad $P(k)$ is a local peak: $I(k) \leftarrow 1$; \\
  8& \qquad \textbf{end} \\
  9& \textbf{end} \\

  10& $F_I \leftarrow $ (the frequency component of $I$ whose element equals 1); \\

  & \textbf{Construct the SIOS:} \\
  11& \textbf{for} $i = 1$ to $length(F_I)$ \\
  12& \qquad $F_c \leftarrow F_I(i)$, $M \leftarrow$ (every element of $F_I$)/$F_c$, $I_N \leftarrow floor(F_c/G)$; \\
  13& \qquad $G(I_G) \leftarrow$ solve inequality \{\textbf{min}$|F_c - I_{N}(k)G(k)| < \Delta_G$\}$|_{k = 1 \; to \; L_G}$; \\

  14& \qquad $\alpha \leftarrow \frac{G(I_G)}{\Delta_G}$, $\beta_{max} \leftarrow floor(\frac{F_s}{2 \times F_c}$); \\

  15& \qquad \textbf{for} $j = 1$ to $\beta_{max}$ \\
  16& \qquad \qquad $\sigma (i,j) \leftarrow j/\alpha$; \\
  17& \qquad \textbf{end} \\
  18& \qquad $M_1 \leftarrow$ (the index of $M$, whose element is integer); \\

  19& \qquad \textbf{for} $k = 1$ to $length(M_1)$ \\
  20& \qquad \qquad $I_m \leftarrow M_1(k)$, $F_{c}^{'} \leftarrow F_I(I_m)$, $\beta \leftarrow M(I_m)$; \\
  21& \qquad \qquad \textbf{if} $\frac{F_{c}^{'}}{G(I_G)} - floor(\frac{F_{c}^{'}}{G(I_G)}) < \sigma (i,\beta)$ \textbf{then} \\
  22& \qquad \qquad \qquad $N(I_G) \leftarrow N(I_G) + 1$; \\
  23& \qquad \qquad \qquad $E(I_G) \leftarrow E(I_G) + P(I_m)$; \\
  24& \qquad \qquad \textbf{end} \\
  25& \qquad \textbf{end} \\
  26& \textbf{end} \\
  \hline
\end{tabular}
\end{center}
\end{table}

\section{Detection of characteristic frequencies of faulty bearings by identifying the dominant frequency}
\subsection{Simulation analysis}
In order to demonstrate the proposed method, we use the similar simulated bearing fault signals that given in \cite{wang2013,wang2011}. The simulated bearing fault signal with one resonant frequency is given as:
\begin{eqnarray}
x_{bear}(k) &=& \sum_{r}\exp^{-\beta\times(k-r\times F_s/f_m-\tau_r)/F_s} \nonumber \\
      && \times\sin(2\pi f\times(k-r\times F_s/f_m-\tau_r)/F_s) \nonumber
\label{sim}
\end{eqnarray}
where $\beta$ is equal to $900$, $f_m$ is the fault characteristic frequency ($f_m=110$Hz), $F_s$ is the sampling frequency ($F_s = 12000$Hz), $\tau_r$ is a uniformly distributed random number which is used to simulate the randomness caused by the slippage, and $f$ is the resonant frequency ($f=3900$Hz). And $k-r\times F_s/f_m-\tau_r \geq 0$ is used to ensure the causality of the exponential function. Gaussian noise is also added and the signal noise ratio is set as $-10$dB.

\begin{figure}[!t]
  \centering
  \includegraphics[width=0.85\textwidth]{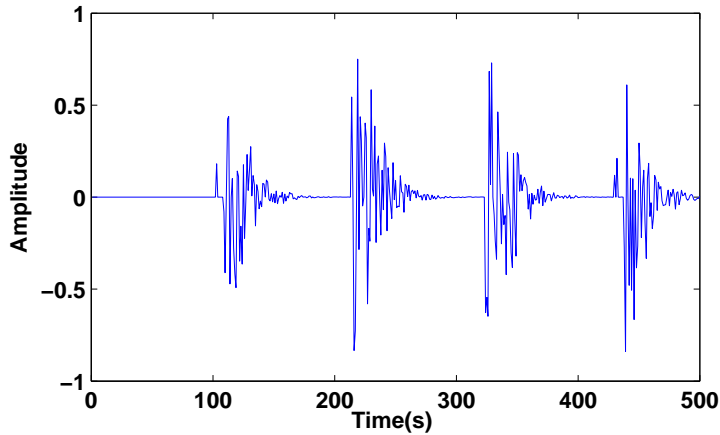}
  \caption{The simulated signal.}
  \label{slip8}
\end{figure}

\begin{figure}[!t]
	\centering
	\includegraphics[width=0.85\textwidth]{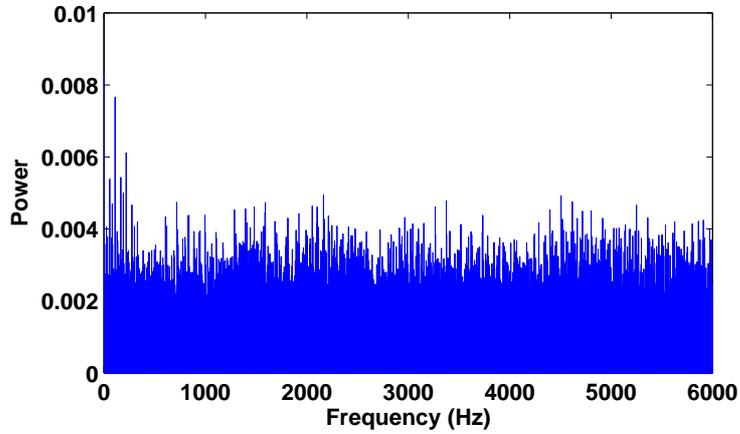}
	\caption{The spectrum of the simulated signal.}
	\label{slip8fft}
\end{figure}

It is known that the slippage of bearings cause smearing of harmonics. In this case the power amplitudes of harmonics will be distributed to adjacent components. Such effects will definitely influence the diagnosis results. Therefore we consider $\tau_r \in [-8,8]$. The original signals are shown in Fig. \ref{slip8}, where only 500 samples are displayed. The corresponding averaged spectra are given in Fig. \ref{slip8fft}.

\begin{figure}[!t]
  \centering
  \includegraphics[width=0.85\textwidth]{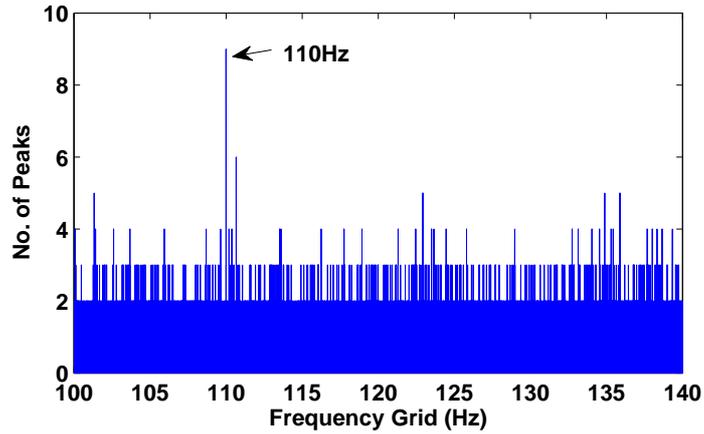}
  \caption{The SIOS of the simulated signal: $N(i)$.}
  \label{slipp8ni}
\end{figure}

\begin{figure}[!t]
  \centering
  \includegraphics[width=0.85\textwidth]{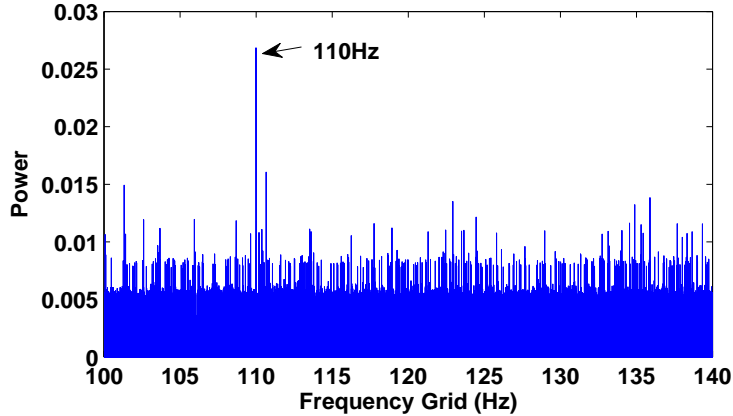}
  \caption{The SIOS of the simulated signal: $E(i)$.}
  \label{slipp8ei}
\end{figure}

The SIOS is given in Fig. \ref{slipp8ni} and Fig. \ref{slipp8ei}, where the frequency grid is selected as $[100\mbox{Hz}, 200\mbox{Hz})$ with $\Delta_G = 0.1 \Delta_s$. The power amplitudes of harmonics are significantly reduced due to smearing as shown in Fig. \ref{slip8fft}. However the proposed method is still effective as shown in Fig. \ref{slipp8ni} and Fig. \ref{slipp8ei}, where $f_m$ is clearly identified. The main reason is that, the SIOS gives information about all harmonics even with small amplitudes through the searching algorithm.

\subsection{Bearing fault signals}
\subsubsection{Case 1: a bearing outer race fault signal obtained from a run-to-failure test}
The vibration signal from a run-to-failure test is used to demonstrate the proposed method. The test rig hosts four test bearings on one shaft driven by an AC motor. The rotation speed is 2000 rpm. A radial load of 6000 lbs. is added to the shaft and bearing. Four Rexnord ZA-2115 double row bearings were installed on one shaft. An Accelerometer was mounted on the housing of each bearing. Vibration data of the bearings collected every 10 min. The data sampling rate is 20 kHz. For more detailed information about this experiment, please refer to \cite{lee2007rexnord}, and the data can be downloaded from Prognostics Center Excellence (PCoE) through prognostic data repository contributed by Intelligent Maintenance System (IMS), University of Cincinnati.

\begin{figure}
	\centering
	\includegraphics[width=0.85\textwidth]{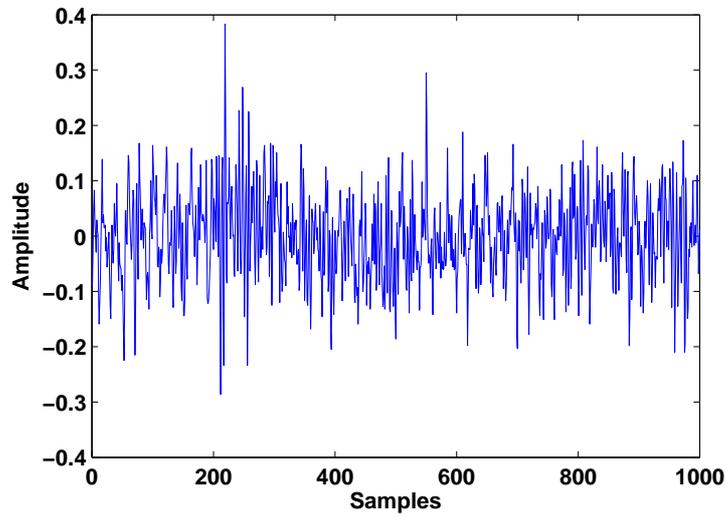}
	\caption{The waveform of record 510.}
	\label{imstime}
\end{figure}

\begin{figure}
	\centering
	\includegraphics[width=0.85\textwidth]{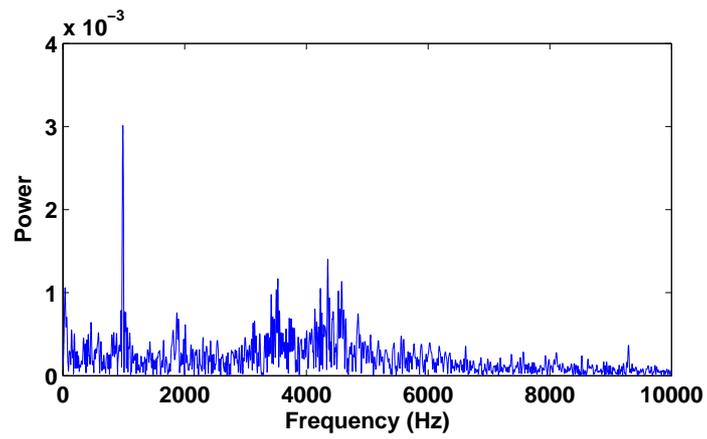}
	\caption{The spectrum of record 510.}
	\label{imsfft}
\end{figure}

The data set collected from February 12nd, 2004 10:32:39 to February 19th, 2004 06:22:39 is used for further analysis. At the end of the test-to-failure experiment, outer race failure occurred in bearing 1. Here we take record 510 which is at February 15, 2004 23:22:39, i.e. the very early stage of the fault. Fig. \ref{imstime} and Fig. \ref{imsfft} show the waveform and the spectrum of record 510.


\begin{figure}
	\centering
	\includegraphics[width=0.85\textwidth]{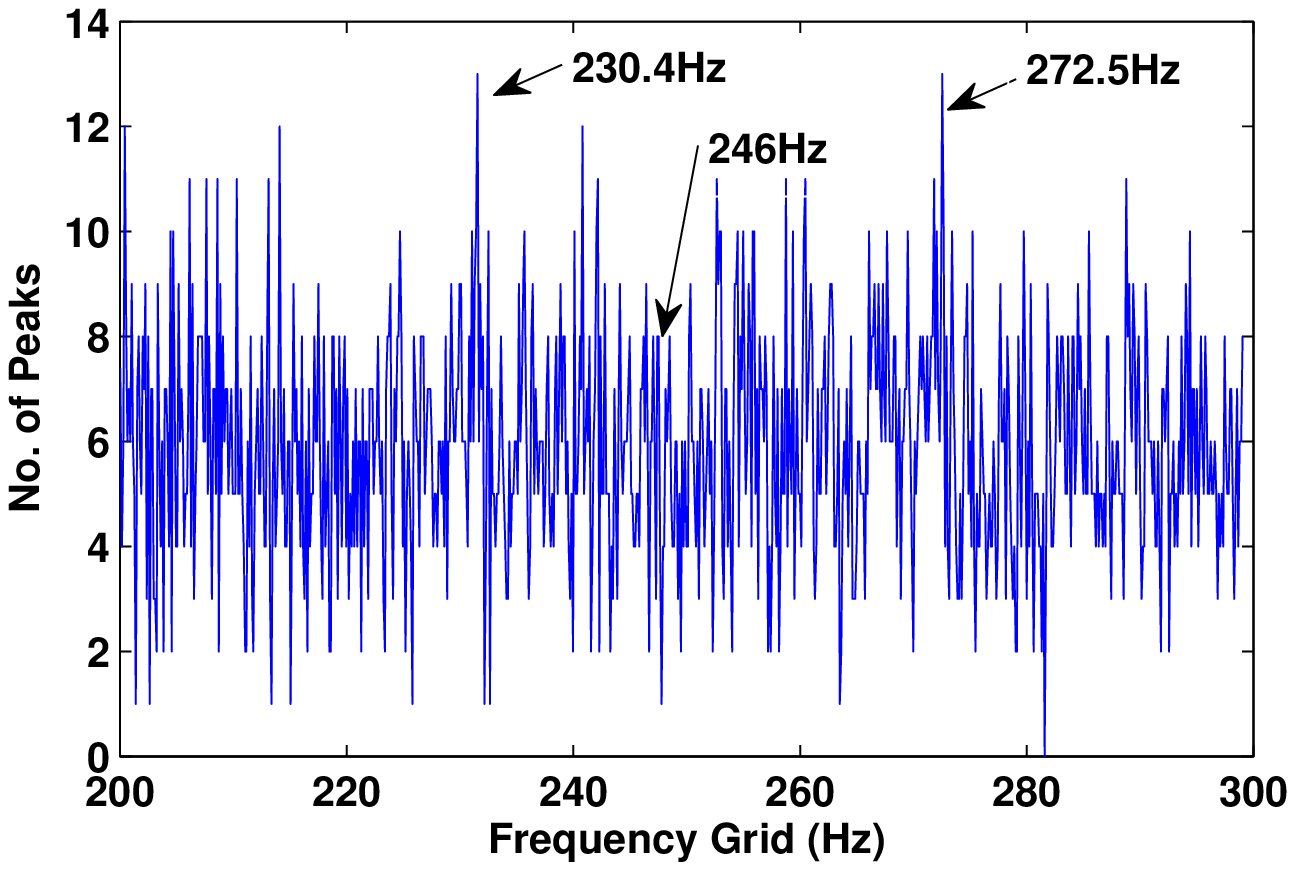}
	\caption{SIOS - $N(i)$ of Record 510.}
	\label{510pni}
\end{figure}

\begin{figure}
	\centering
	\includegraphics[width=0.85\textwidth]{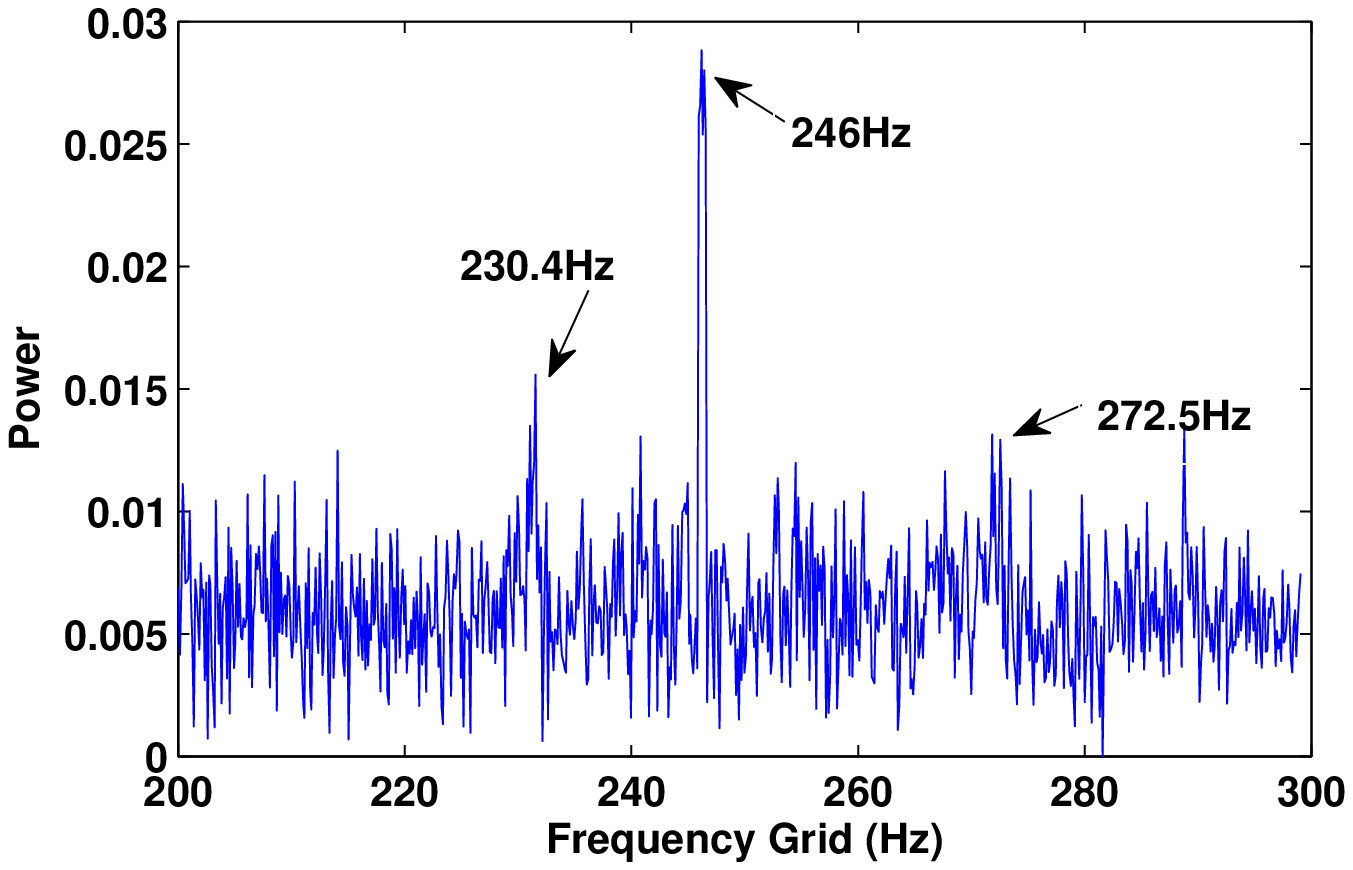}
	\caption{SIOS - $E(i)$ of Record 510.}
	\label{510pei}
\end{figure}

The frequency grid is selected as $[200\mbox{Hz}, 300\mbox{Hz})$, and $\Delta_G = 0.1 \Delta_s$. Fig. \ref{510pni} and \ref{510pei} depict the SIOS of record 510 with outer-race fault. Although $246$Hz is a dominant in Fig. \ref{510pei}, the number of peaks found around $246$Hz is quite small as shown in Fig. \ref{510pni}. Hence $246$Hz is not treated as the dominant component on the frequency grid. The similar is $272.5$Hz, which has a smaller power than $230.4$Hz. Only $230.4$Hz is significant in both figures, and therefore it is identified as the dominant component on the frequency grid. It is clearly corresponding to the outer-race fault \cite{zhang2015weak}.

\begin{figure}
	\centering
	\includegraphics[width=0.85\textwidth]{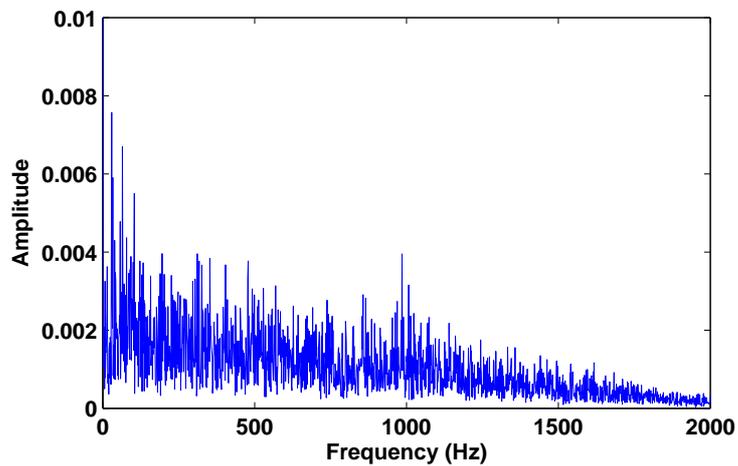}
	\caption{The envelope spectrum of filtered signal of Record 510. The optimal band selected through spectral kurtosis method is $[2800,5100]$.}
	\label{sk510}
\end{figure}

As demonstrated by the results, the outer-race fault is detected after around 3.54 running days. We also apply spectral kurtosis based method to find the characteristic frequency, and the envelope spectrum of filtered signal is shown in Fig. \ref{sk510}, in which no evident characteristic frequency can be observed. The proposed method also detects the fault earlier than the method in \cite{zhang2015weak}, where the fault is detected after around 3.8 running days.


\subsubsection{Case 2: a rolling element bearing benchmark}
The vibration data from the Case Western Reserve University (CRWU) Bearing Data Center \cite{bearingdata} are analyzed and compared with the published benchmark results in \cite{smith2015}. The test stand consists of a 2 hp motor (left), a torque transducer/encoder (center), a dynamometer (right), and control electronics. The test bearings support the motor shaft. With the help of electrostatic discharge machining, inner-race and outer-race faults of different sizes are made. The vibration data are collected using accelerometers attached to the housing with magnetic bases. In this study, the driver end (DE) data with the sampling frequency being 12000 Hz are analyzed. The characteristic frequencies of bearings, i.e. ball pass frequency of outer-race (BPFO), ball pass frequency of inner-race (BPFI), fundamental train frequency (FTF) and ball spin frequency (BSF), are shown in Table \ref{tab1}. And we use $f_r$ to denote the rotating frequency.
\begin{table}[!t]\small
\renewcommand{\arraystretch}{1.3}
\begin{center}
\caption{Bearing fault frequencies (multiple of running speed in Hz)}\label{tab1}
\begin{tabular}{cccc}
  \hline
 BPFO & BPFI & FTF & BSF\\
 \hline
 3.585 $\times f_r$  & 5.415 $\times f_r$  &  0.3983 $\times f_r$ & 2.357 $\times f_r$\\
 \hline
\end{tabular}
\end{center}
\end{table}

The frequency grid is selected as $[100\mbox{Hz}, 180\mbox{Hz})$. $\Delta_G = 0.1 \Delta_s$ and $l = 10000$ are chosen for all records. $\delta$ is set as 0.0002 for most records, where $\delta$ is set as 0.002 for record 3005-3008.

We firstly take records 105, 130, 118 as examples to demonstrate the results, which are with inner-race fault, outer-race fault and ball fault respectively.
\begin{figure}[!t]
  \centering
  \includegraphics[width=0.85\textwidth]{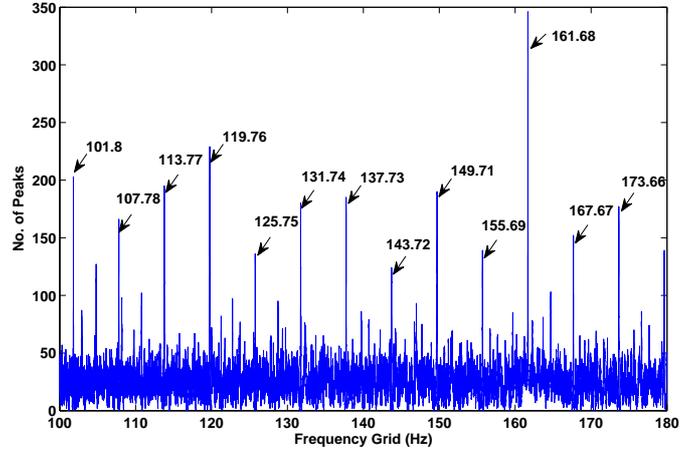}
  \caption{SIOS - $N(i)$ of Record 105 (1797rpm), inner-race fault.}
  \label{105ni}
\end{figure}
\begin{figure}[!t]
  \centering
  \includegraphics[width=0.85\textwidth]{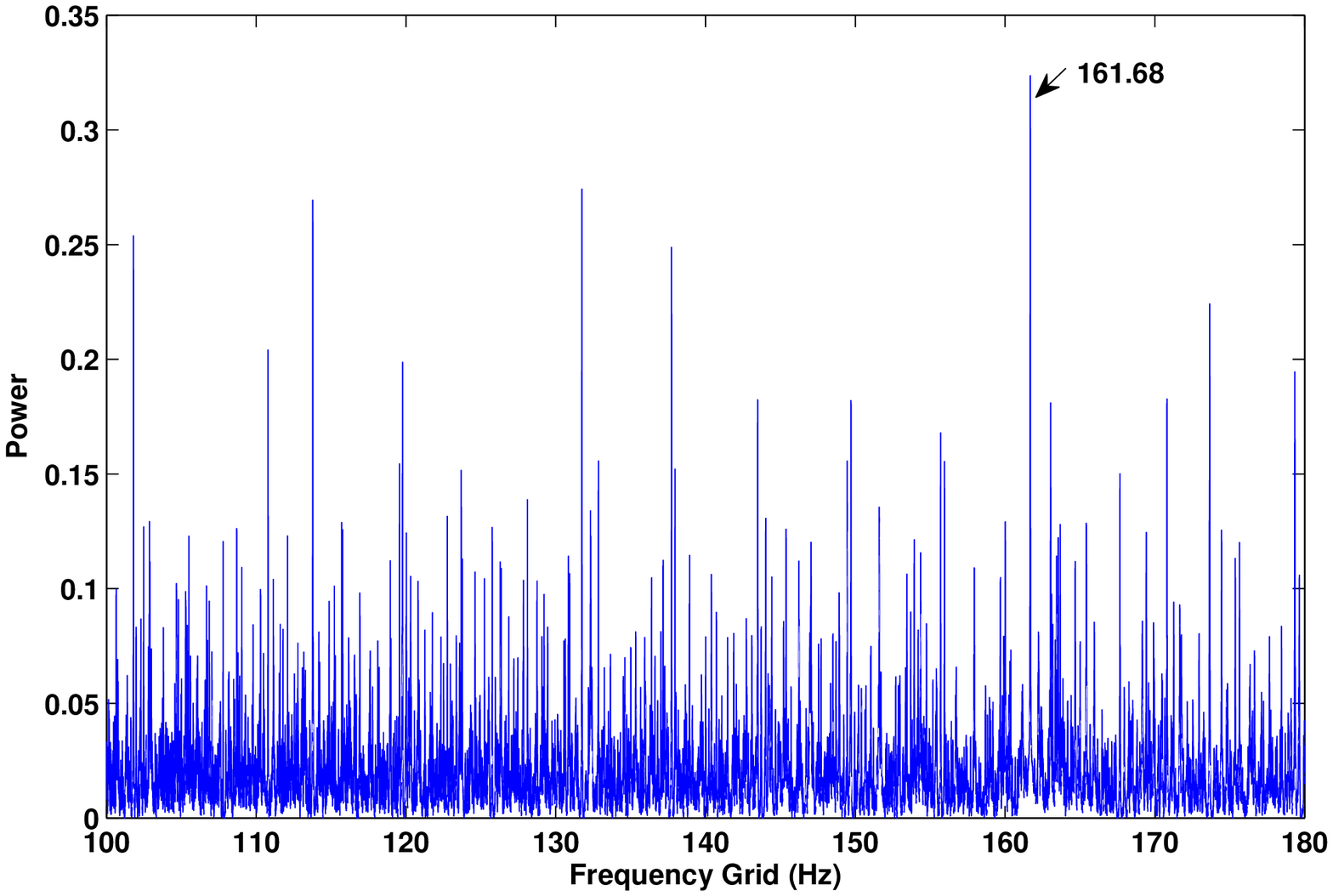}
  \caption{SIOS - $E(i)$ of Record 105 (1797rpm), inner-race fault.}
  \label{105ei}
\end{figure}
\begin{figure}[!t]
  \centering
  \includegraphics[width=0.85\textwidth]{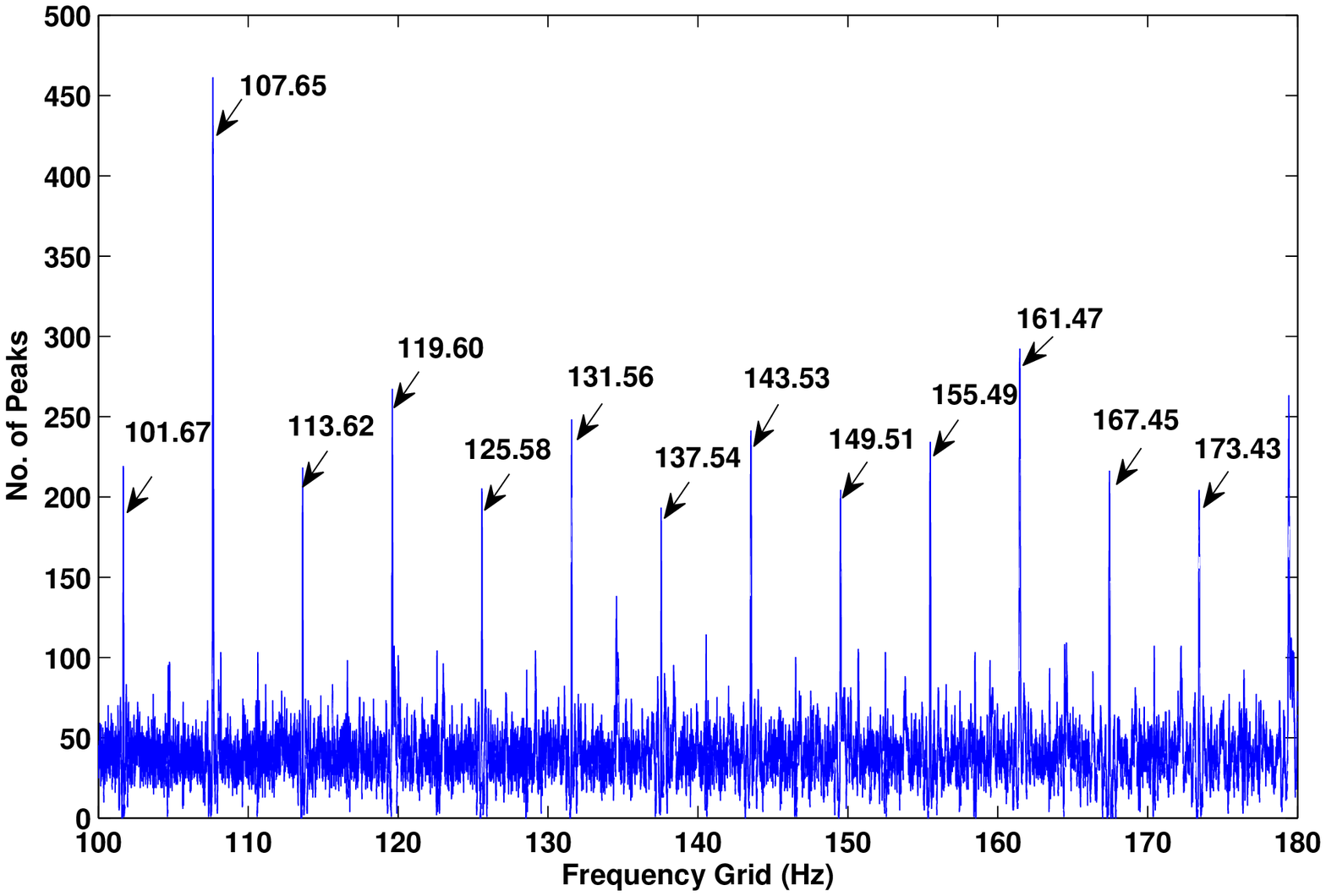}
  \caption{SIOS - $N(i)$ of Record 130 (1797rpm), outer-race fault.}
  \label{130ni}
\end{figure}
\begin{figure}[!t]
  \centering
  \includegraphics[width=0.85\textwidth]{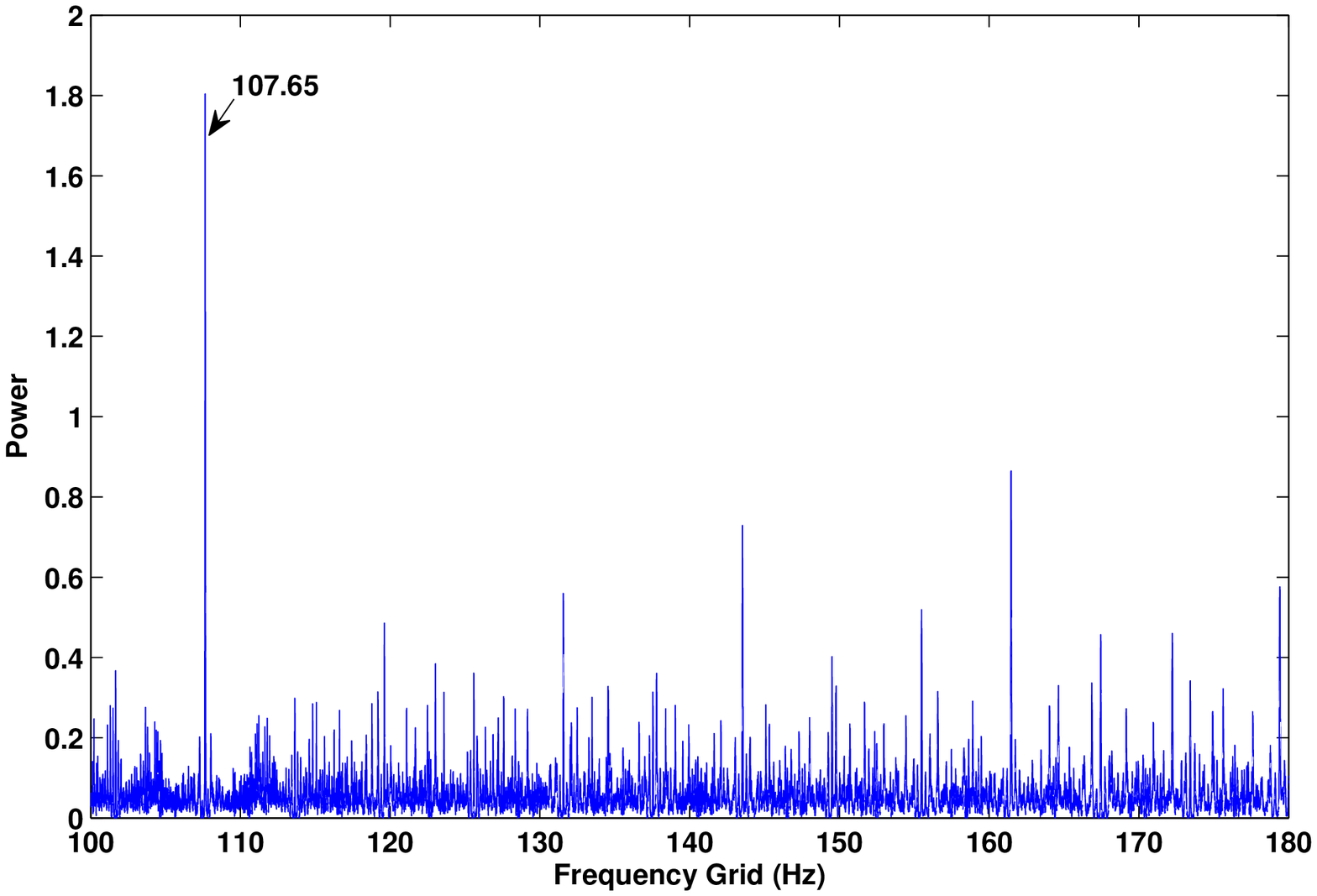}
  \caption{SIOS - $E(i)$ of Record 130 (1797rpm), outer-race fault.}
  \label{130ei}
\end{figure}
\begin{figure}[!t]
  \centering
  \includegraphics[width=0.85\textwidth]{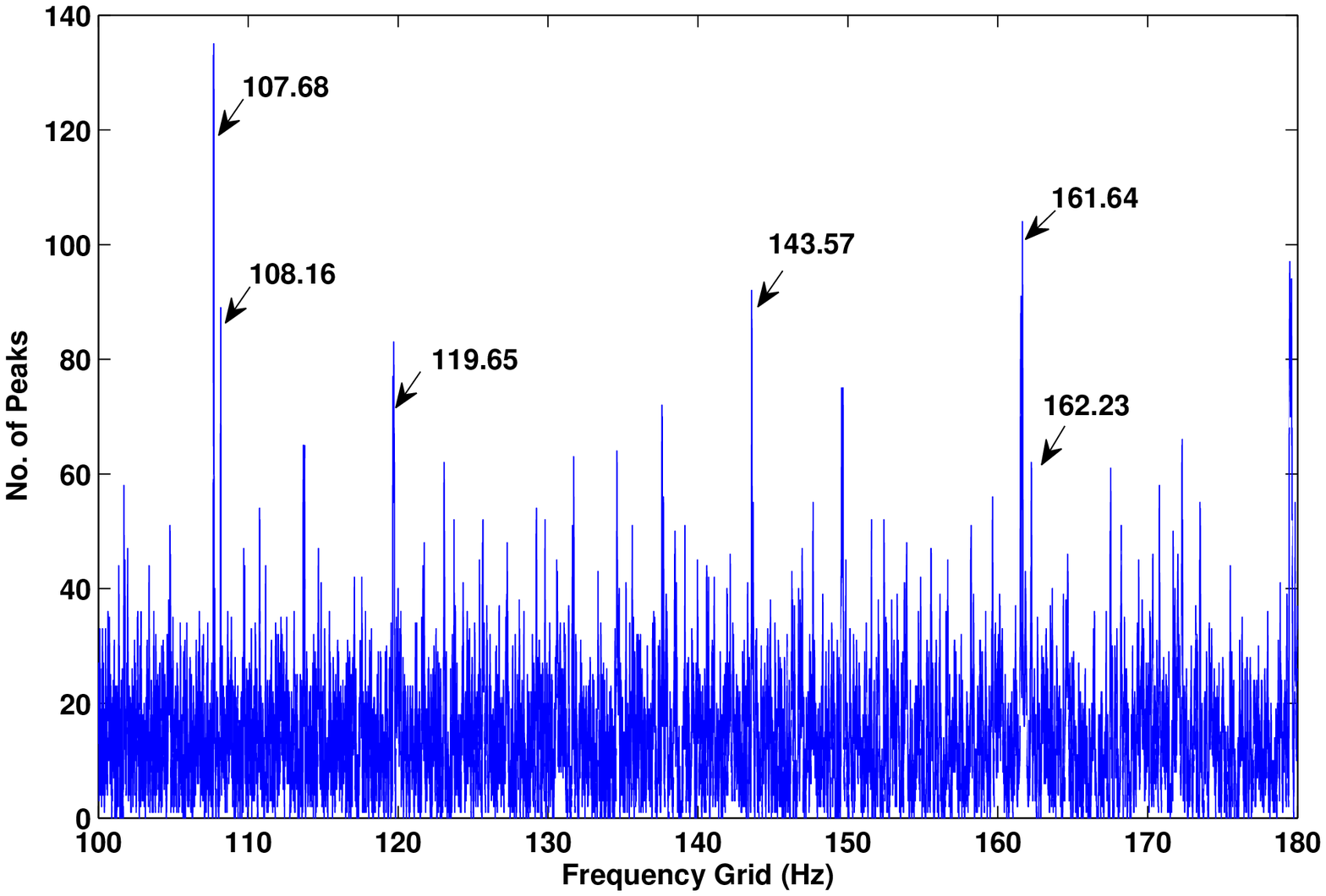}
  \caption{SIOS - $N(i)$ of Record 118 (1797rpm), ball fault.}
  \label{118ni}
\end{figure}
\begin{figure}[!t]
  \centering
  \includegraphics[width=0.85\textwidth]{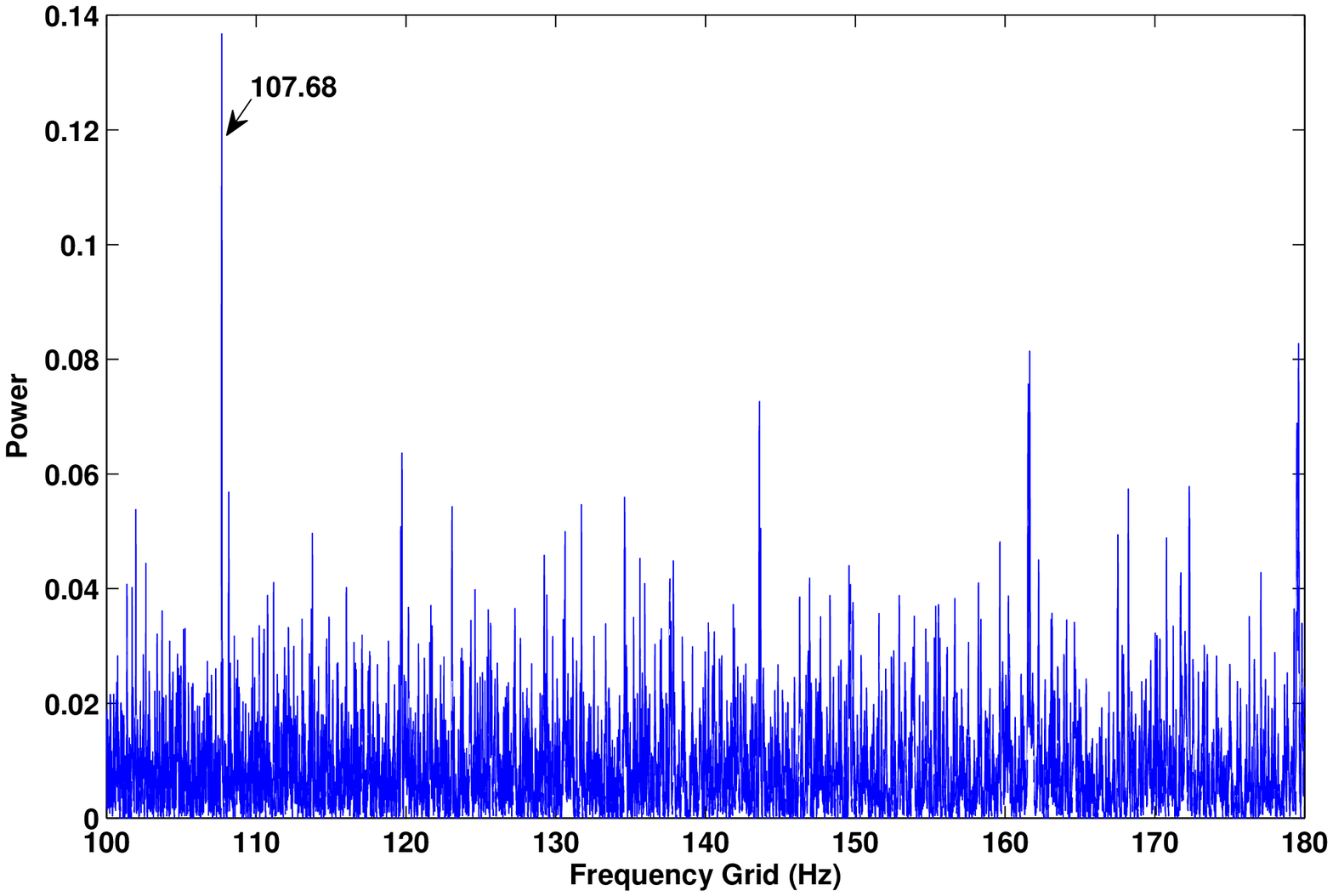}
  \caption{SIOS - $E(i)$ of Record 118 (1797rpm), ball fault.}
  \label{118ei}
\end{figure}

Fig. \ref{105ni} and \ref{105ei} depict the SIOS of record 105 with inner-race fault. The dominant component can be easily identified on the frequency grid, which is $161.68$Hz. It is clearly corresponding to the inner-race fault ($1\times$ BPFI). Fig. \ref{130ni} and \ref{130ei} depict the SIOS of record 130 with outer-race fault. The dominant component can also be easily identified on the frequency grid, which is $107.65$Hz. It is clearly corresponding to the outer-race fault ($1\times$ BPFO).

It is interesting that, all significant components in Fig. \ref{105ni} and \ref{130ni} seem to be harmonics of $0.2\times f_r$. Similar observations have been found in \cite{smith2015} when some records with ball fault were analyzed with the envelope spectrum. As stated in \cite{smith2015}, it is quite likely that the amount of mean slip in bearing has adjusted itself to lock onto an exact subharmonic of a dominant frequency such as shaft speed. In this case BPFI seems to lock onto $5.4\times f_r$ and BPFO seems to lock onto $3.6\times f_r$, which are both multipliers of $0.2\times f_r$. This is also confirmed by most of records with inner-race fault or outer-race fault, whose harmonics of $0.2\times f_r$ are significant on the SIOS and the component corresponding to $5.4\times f_r$ or $3.6\times f_r$ is the dominant one in $N(i)$ and $E(i)$ of the SIOS.

Fig. \ref{118ni} and \ref{118ei} depict the SIOS of record 118 with ball fault. An interesting feature is that the ball fault records often present evidence of outer and inner race faults, which was also discussed in \cite{smith2015}. Most of the marked components in Fig. \ref{118ni} are harmonics of $0.2\times f_r$, among which $107.68$Hz is related to $3.6\times f_r$ ($1\times$ BPFO) and $143.57$Hz is related to $2.4\times f_r$ ($1\times$ BSF) and $119.65$ is related to $0.4\times f_r$ ($1\times$ FTF). Nevertheless the components with $162.23$Hz and $108.16$Hz are harmonics of $0.2006\times f_r$ other than $0.2\times f_r$. The component with $162.23$Hz is related to $5.416\times f_r$ and the component with $108.16$ is related to $3.611\times f_r$. It is still difficult to explain why the harmonics of $0.2006\times f_r$ and $0.2\times f_r$ exist at the same.

It is worth mentioning the ball fault pattern (BFP) composed of $3.6\times f_r$, $3.611 \times f_r$, $5.4\times f_r$ and $5.416\times f_r$ (like 107.68Hz, 108.16Hz, 161.64Hz and 162.23Hz in Fig. \ref{118ni}) exists in records 118-121, 185-188 and 222-225.

Based on the above analysis, the bearings are diagnosed based on the SIOS according to the following rules:
\begin{itemize}
  \item inner-race fault: the dominant component in SIOS is BPFI, 
  \item outer-race fault: the dominant component in SIOS is BPFO, and 
  \item ball fault: the BFP, $2\times$BSF, and at least one harmonic of FTF are significant in SIOS; or $2\times$ BSF is dominant in SIOS.
\end{itemize}
If one of the above rules is fully satisfied, the diagnose result is marked by Y; and if one of the above rules is partly satisfied, the diagnose result is marked by P. If no rules is satisfied, the diagnose result is marked by N.

\begin{figure}[!t]
  \centering
  \includegraphics[width=0.85\textwidth]{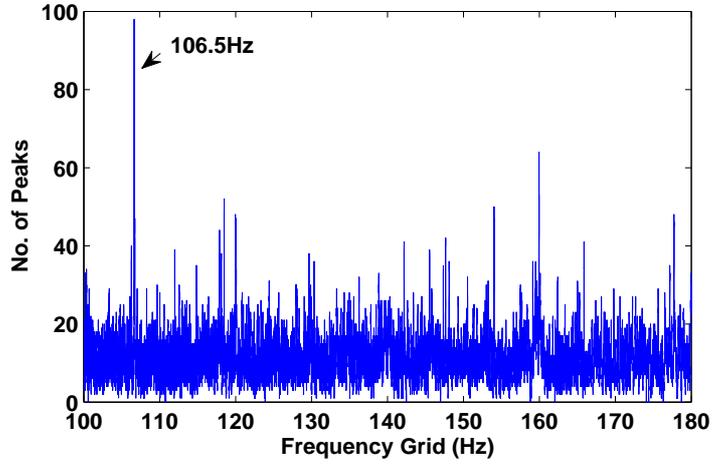}
  \caption{SIOS - $N(i)$ of Record 198 (1772rpm), outer-race fault.}
  \label{198ni}
\end{figure}

\begin{figure}[!t]
  \centering
  \includegraphics[width=0.85\textwidth]{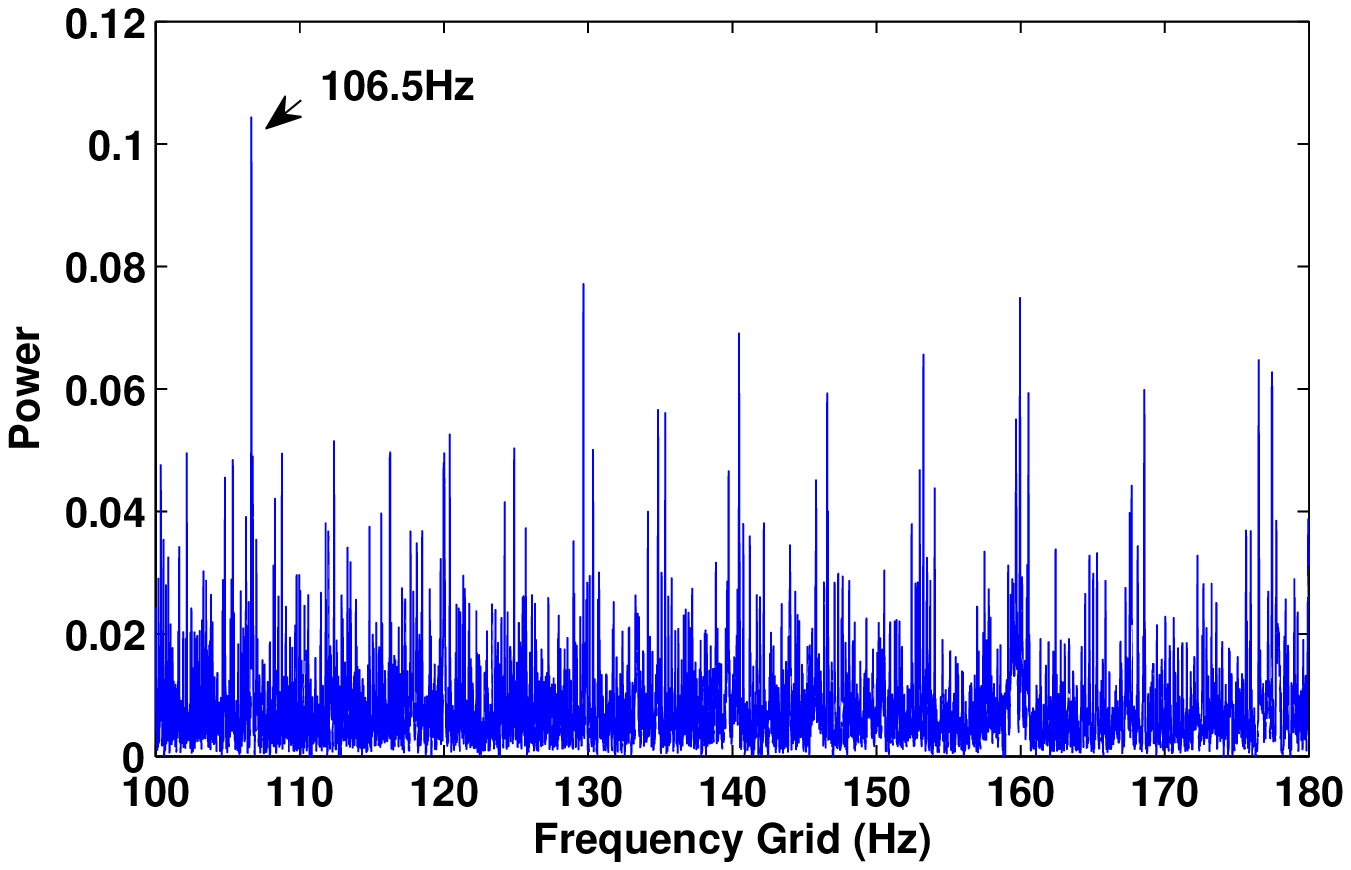}
  \caption{SIOS - $E(i)$ of Record 198 (1772rpm), outer-race fault.}
  \label{198ei}
\end{figure}

\begin{figure}[!t]
	\centering
	\includegraphics[width=0.85\textwidth]{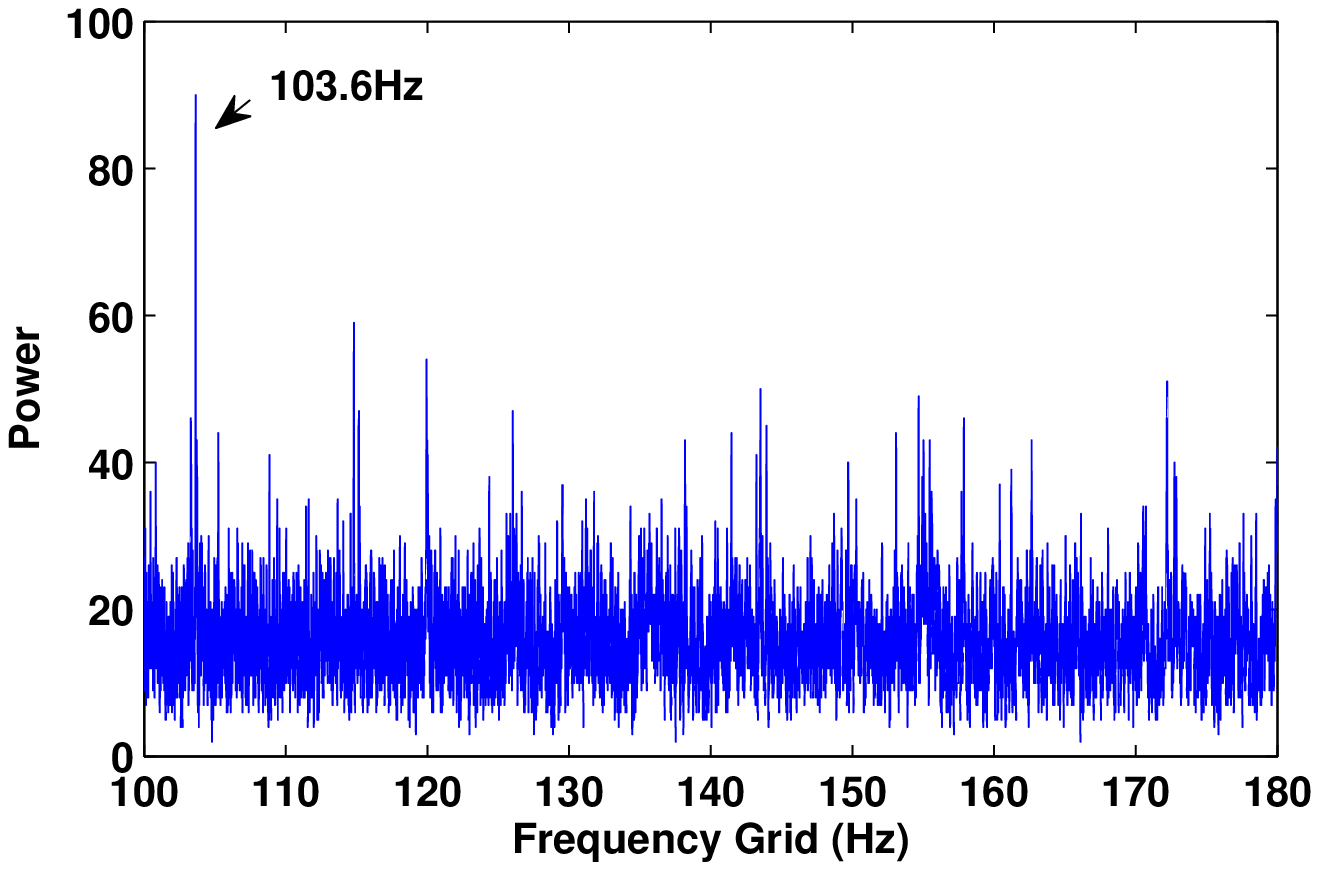}
	\caption{SIOS - $N(i)$ of Record 200 (1730rpm), outer-race fault.}
	\label{200ni}
\end{figure}

\begin{figure}[!t]
	\centering
	\includegraphics[width=0.85\textwidth]{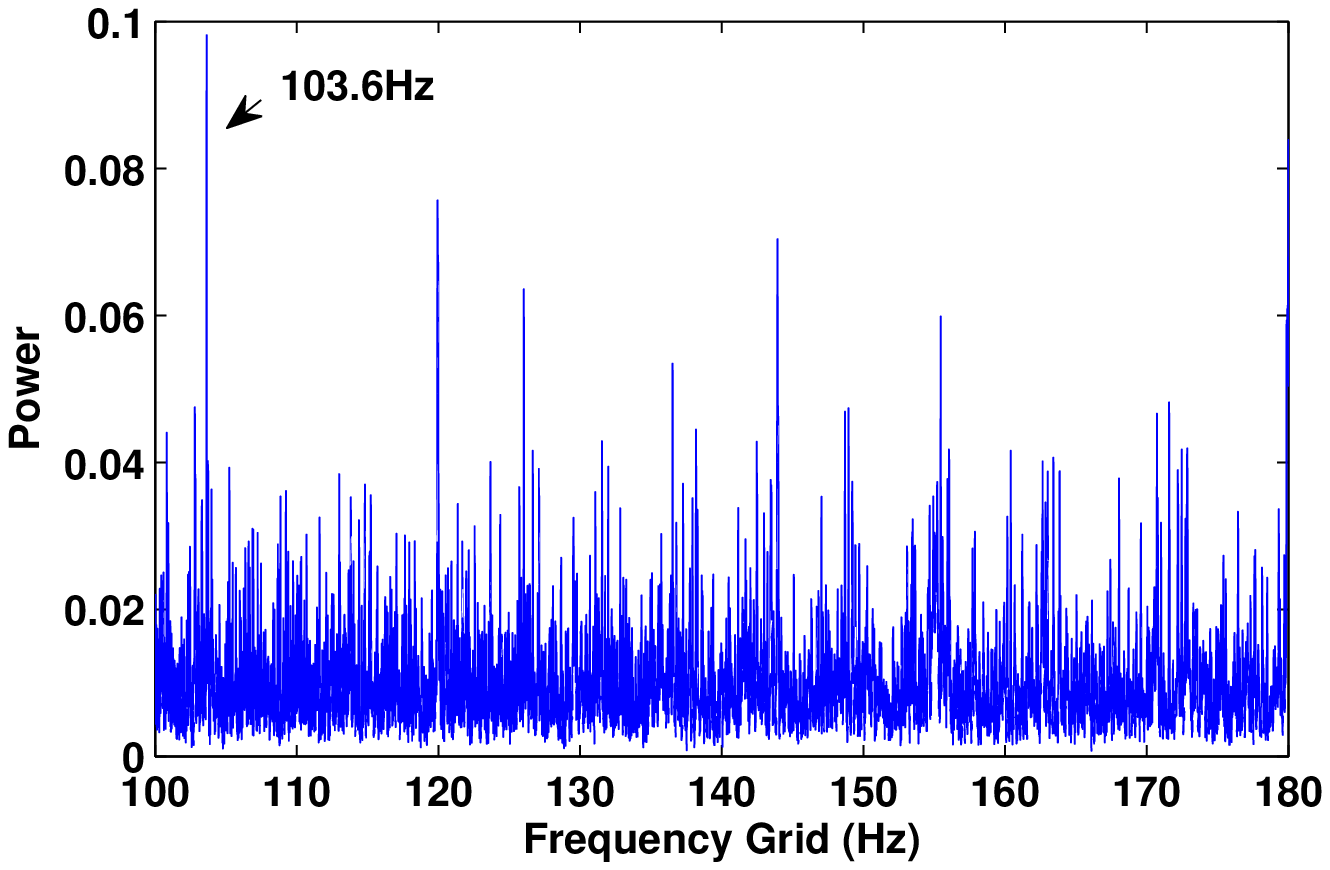}
	\caption{SIOS - $E(i)$ of Record 200 (1730rpm), outer-race fault.}
	\label{200ei}
\end{figure}

The diagnosis results are compared with the benchmark study in \cite{smith2015}, where envelope analysis, spectral kurtosis and cepstrum techniques were applied. As listed in Table B2 given by \cite{smith2015}, record 198, 199, 200 with outer-race fault can not be diagnosed or only partly diagnosed through the benchmark method. We take record 198 and 200 to illustrate the effectiveness of the proposed method as shown in Fig. \ref{198ni} - \ref{200ei}. It is clear that, BPFO is the dominant component in SIOS, and therefore the outer-race fault can be successfully diagnosed by proposed method.

Besides, most of records with ball fault were marked by N1 in \cite{smith2015}, because it was difficult to explain why some BSF components in the envelope spectrum were much stronger than the others. In fact the stronger BSF components in the envelop spectrum often coincide with the BPFI or BPFO components. In other words, once the BSF component conform with a BPFI or BPFO component, it could be a strong one. That is also the reason why we observe significant components corresponding to BPFI and BPFO in SIOS of records with ball faults (see Fig. \ref{118ni}). In this sense the diagnose results of those records with ball faults can also be marked by Y or P according to the rule defined in \cite{smith2015}. We will not compare those records due to the ambiguous results of the benchmark.

Table \ref{tab2} - Table \ref{tab4} give the full diagnosis results with the proposed method and the comparison with benchmark. Record 121 and 188 with ball fault can not be fully diagnosed by the proposed method, where the BFP can not be fully identified and $2\times$ BSF is not dominant in the SIOS. With the proposed method satisfied results can also not be obtained for record 3001-3004, and the reason is similar with that demonstrated in \cite{smith2015}.


\begin{table}[!t]\small
\renewcommand{\arraystretch}{1.3}
\begin{center}
\caption{12K drive end bearing fault analysis results: Inner-race fault. DE only. Y=successful, P=partially successful, N=not successful}\label{tab2}
\begin{tabular}{ccc}
  \hline
 Inner-race faults & & \\
 \hline
 Data Set & Diagnosis Result  &  Benchmark Result (M1,M2,M3)\\
 \hline
 105 & Y  & Y,-,- \\
 106 & Y & Y,-,- \\
 107 & Y & Y,-,- \\
 108 & Y & Y,-,- \\
 169 & Y & Y,Y,Y \\
 170 &Y & Y,Y,Y \\
 171 &Y & Y,Y,Y \\
 172 &Y & Y,Y,Y \\
 209 &Y & Y,-,- \\
 210 &Y & Y,-,- \\
 211 &Y &  Y,-,- \\
 212 &Y &  Y,-,- \\
 3001 & N & N,N,N \\
 3002 &N& N,N,N \\
 3003 &N& N,N,N \\
 3004 &N& N,N,N \\
  \hline
\end{tabular}
\end{center}
\end{table}

\begin{table}[!t]\small
\renewcommand{\arraystretch}{1.3}
\begin{center}
\caption{12K drive end bearing fault analysis results: Outer-race fault. DE only. Y=successful, P=partially successful, N=not successful}\label{tab3}
\begin{tabular}{ccc}
  \hline
 Outer-race faults & \\
  \hline
 Data Set & Diagnosis Result  &  Benchmark Result (M1,M2,M3)\\
 \hline
 130&Y & Y,-,- \\
 131&Y & Y,-,- \\
 132&Y & Y,-,- \\
 133&Y & Y,-,- \\
  197&Y & N,N,Y \\
 198&Y & P,N,N \\
 199&Y & P,N,N \\
 200&Y & N,N,N \\
  234&Y & Y,-,- \\
 235&Y& Y,-,- \\
 236&Y & Y,-,- \\
 237&Y & Y,-,- \\
 \hline
\end{tabular}
\end{center}
\end{table}

\begin{table}[!t]\small
\renewcommand{\arraystretch}{1.3}
\begin{center}
\caption{12K drive end bearing fault analysis results: Ball fault. DE only. Y=successful, P=partially successful, N=not successful}\label{tab4}
\begin{tabular}{ccc}
  \hline
 Ball faults & \\
  \hline
 Data Set & Diagnosis Result  &  Benchmark Result (M1,M2,M3)\\
 \hline
 118&Y & -,-,- \\
 119&Y & -,-,- \\
 120&Y & -,-,- \\
 121&P & -,-,Y \\
 185&Y & P,P,P \\
 186&Y & P,P,P \\
 187&Y & P,-,P \\
 188&P & P,P,P \\
 222&Y & P,Y,Y \\
 223&Y & Y,Y,Y \\
 224&Y & -,-,P \\
 225&Y & -,-,P \\
 \hline
\end{tabular}
\end{center}
\end{table}

\section{Discussion}
\subsection{About the selection of $l$ and $\delta$ in (\ref{localJ})}
According to (\ref{localJ}), the number of local peaks is determined by the selection of $l$ and $\delta$. Since the average power in different frequency bands may be varying, we use $l$ in (\ref{localJ}) to build a frequency-dependent baseline in order to suppress frequency components with low power amplitudes. The selection of $l$ is not strict, and it can be simply chosen such that $2l+1$ components represent any desired frequency bandwidth, e.g. 50Hz, 100Hz. Then the number of local peaks will be controlled through the selection of $\delta$. If $\delta$ is too large, then only a few local peaks can be found and therefore some harmonics may be lost. If $\delta$ is too small, the number of local peaks could be large and the searching effort could be very high. Since the harmonics of bearing characteristic frequencies usually have relatively larger amplitudes, in practice we suggest to select $\delta$ such that $0.5\% \thicksim 3\%$ of amplitudes are treated as local peaks.

In the benchmark study $l$ is set as 10000, such that the bandwidth of 114Hz is used to compute the moving average. And $\delta$ is set as 0.0002 for most records except record 3005-3008. Since the spectra of different loads and different faults (with different sizes) are quite different, in fact $0.5\% \thicksim 2.1\%$ of amplitudes are treated as local peaks in the SIOS of different spectra with the same $l$ and $\delta$. From this point of view, the use of the same parameters for those different spectra have already demonstrated that $l$ and $\delta$ are not sensitive to the construction of the SIOS if they are within a reasonable range.

\subsection{About the determination of the frequency grid}
If the characteristic frequencies of a given type of bearings are roughly estimated according to the geometrical parameters, the frequency grid can be selected according to the range of estimated characteristic frequencies. If the characteristic frequencies are completely unknown, the range of the frequency grid could be the same with that of the spectrum. In this way the harmonics of all components of the original spectrum are considered in the SISO.

Indeed there is no any restriction on the selection of $G$. The purpose of selecting $F_l$ and $F_h$ in (\ref{basis}) is to reduce the computation effort of the searching.

\subsection{Advantages and limitations}
The proposed method is based on a simple searching algorithm, and it is effective in finding the harmonics of the interested frequency range.  Even the harmonics with small amplitudes could be found and projected onto the frequency grid. Although noises and other random impulses may introduce some local peaks unrelated to the faults, the significant components in the SIOS can still be clearly recognized and related to the characteristic frequencies of bearings.

The proposed method is robust against noises and random impulses. Similar methods, e.g. cepstrum, usually could not find enough periodic components of the signal in case of heavy noises and other interferences. As illustrated in the benchmark study of bearings, the proposed method could provide more information about the harmonics of impulses than the envelope analysis and better diagnosis results can be achieved.

Limitations of the proposed method are as follows: (a) the SIOS can only provide qualitative information, where $N(i)$ and $E(i)$ are not the true value due to the finite spectrum resolution and noises; (b) the computation effort could be high, when the range of the frequency grid or the number of local peaks is large; and (c) the method can not work if the fault has no signatures made of peaks.

\section{Conclusion}
A simple and effective method for detecting bearing faults has been proposed based on a searching algorithm. The SIOS of vibration signals was constructed, such that the information of the train of harmonics was clearly represented with $N(i)$ and $E(i)$ on a defined frequency grid. The dominant or significant components of the SIOS were corresponding to the characteristic frequency of bearings. Based on the SIOS faults of bearings were successfully diagnosed. Its effectiveness was verified with simulated bearing signals and the experimental results.

\section*{Acknowledgements}
The research is supported by National Natural Science Foundation of China (grant number 51475455), and the Project Funded by the Priority Academic Program Development of Jiangsu Higher Education Institutions(PAPD).

\bibliography{MST-Mybibfile_V1}

\end{document}